\documentclass[reprint,aip]{revtex4-2}

\usepackage[version=4]{mhchem}
\usepackage{siunitx}
\DeclareSIUnit\Molar{M}

\usepackage{times}
\usepackage[dvipsnames]{xcolor}
\usepackage{tikz}
\usepackage[export]{adjustbox}
\usepackage{graphicx}% Include figure files
\usepackage{mathtools}
\usepackage{epsfig}
\usepackage{epstopdf}
\usepackage{textcomp}
\usepackage{color,sidecap,fancyhdr,ulem,xcolor}
\usepackage{textcomp}
\usepackage{nicefrac}
\usepackage{xfrac}
\usepackage{amsmath}
\usepackage{amssymb,amsthm,wrapfig,esint,afterpage,marginnote,afterpage,lipsum,tabularx}
\usepackage{mathpazo}
\usepackage[utf8]{inputenc}
\usepackage{natbib}

\newcommand{\pb}{\mathbf{P}}

\definecolor{ao(english)}{rgb}{0.0, 0.5, 0.0}

\begin{document}
	
	% A title suggested by Leah and I like it
	\title{A mass conserved reaction-diffusion system reveals switching between coexisting polar and oscillatory cell motility states}
	
	% Alternative option
	%\title{Switching between coexisting polar and oscillatory cell motility}
	
	\author{Jack M. Hughes}
	\affiliation{Department of Mathematics, University of British Columbia, Vancouver, Canada}
	\author{Cristina Martinez-Torres}
	\affiliation{Institute of Physics and Astronomy, University of Potsdam, Potsdam, 14476, Germany}
	\author{Carsten Beta}
	\affiliation{Institute of Physics and Astronomy, University of Potsdam, Potsdam, 14476, Germany}
	\author{Leah Edelstein-Keshet}
	\affiliation{Department of Mathematics, University of British Columbia, Vancouver, Canada}
	\author{Arik Yochelis}
	\affiliation{Swiss Institute for Dryland Environmental and Energy Research, Sede Boqer Campus, Midreshet Ben-Gurion 8499000}
	\affiliation{Department of Physics, Ben-Gurion University of the Negev, Be'er Sheva 8410501, Israel} \email{yochelis@bgu.ac.il}
	
	\date{\today}
	
	\begin{abstract}
		Motile eukaryotic cells display distinct modes of migration that often occur within the same cell type. It remains unclear, however, whether transitions between the migratory modes require changes in external conditions, or whether the different modes are coexisting states that emerge from the underlying signaling network. Using a mass-conserved reaction-diffusion model of small GTPase signaling with F-actin mediated feedback, we uncover a bistable regime where a polarized mode of migration coexists with spatiotemporal oscillations. Indeed, experimental observations of \textit{D.~discoideum} show that, upon collision with a rigid boundary, cells may switch from polarized to disordered motion.
		% 600 chars
	\end{abstract}
	\maketitle
	
	\noindent \textbf{Eukaryotic cells display distinct migration modes. While some migratory modes are tightly linked to specific functions and cell types, also migratory plasticity may occur, where different modes of locomotion are observed within the same cell type. It remains unclear, however, whether transitions between the migratory modes require changes in external conditions, or whether the different modes are coexisting states that emerge from the underlying signaling network. This theoretical and experimental study provides a distinct approach for addressing the question of internally coexisting dynamical structures in cell motility and presents the first potential mechanism of this essential aspect of cortical pattern formation.} \\ \\
	
	Motile eukaryotic cells play a key role in many important biological processes, such as early embryonic development or functions of the immune system~\cite{sengupta_principles_2021,yamada_mechanisms_2019}. The mechanical forces that drive their locomotion are generated by the assembly of filamentous actin (F-actin) that causes protrusion of the cell edge~\cite{bray2000cell}.
	The cytoskeletal dynamics is governed by upstream signaling pathways and provides the basis not only for motility but also for other central cellular functions, such as nutrient uptake and cell division~\cite{bement2015activator,xiao_mitotic_2017,flemming_how_2020,veltman2016plasma,bernitt2015dynamics,bernitt2017fronts,buccione2004foot}. Also numerous pathological phenomena are related to defects in F-actin regulation, among these, aberrant cell motility in metastatic cancer~\cite{stuelten_cell_2018}.
	
	Actin-driven cell motility relies on self-organized space-time patterns that emerge from the underlying regulatory dynamics~\cite{Shibata2012,westendorf_actin_2013,Lomakin2015,negrete_noisy_2016,diz2016membrane,rappel_mechanisms_2017,devreotes_excitable_2017,alonso2018modeling,Horning2019,stankevicins_deterministic_2020,liu2021spots,ecker2021excitable}.
	Central players in this regulatory network are small GTPases that interact rapidly to form a biochemical ``pre-pattern'' inside the cell.
	Small GTPases have been intensely studied experimentally~\cite{jaffe2005rho,ridley2001rho} as well as theoretically over the past decades~\cite{jilkine2007mathematical,otsuji2007mass,Mori2008,mori2011asymptotic,jilkine2011comparison,rubinstein2012weakly,verschueren2017model,liu2021spots}.
	While small GTPases may locally promote nucleation and polymerization of F-actin, they also interact with F-actin via feedback loops.
	This gives rise to a variety of space-time patterns broadly known as ``actin waves''~\cite{vicker_f_actin_2002,giannone2004periodic,dobereiner2006lateral,gerisch2009self,barnhart2017adhesion,iwamoto2025excitable}, a prominent example of pattern formation at the subcellular level, see~\cite{beta2023actin,noguchi2025nonequilibrium} and the references therein.
	
	Eukaryotic cells can move in many different ways, ranging from disordered (random-like walks) to highly persistent migration~\cite{petrie_random_2009,sengupta_principles_2021}.
	The different modes of locomotion are related to distinct cell shapes and cytoskeletal arrangements, such as the flat, extended, and stable leading edge (lamellipodium) of fish keratocytes or the small, compact, and short-lived membrane protrusions (pseudopodia) of amoeboid cells.
	While some migratory modes are tightly linked to specific functions and cell types, migratory plasticity can also occur, where distinct modes of locomotion occur within the same cell type.
	For example, during cancer progression, metastatic cells undergo transitions between different migratory modes, such as amoeboid and mesenchymal motility~\cite{friedl_cancer_2011,nikolaou_stressful_2020}.
	
	%***********************************************************************
	\begin{figure*}[tp]
		%\centering
		\includegraphics[width=0.9\textwidth]{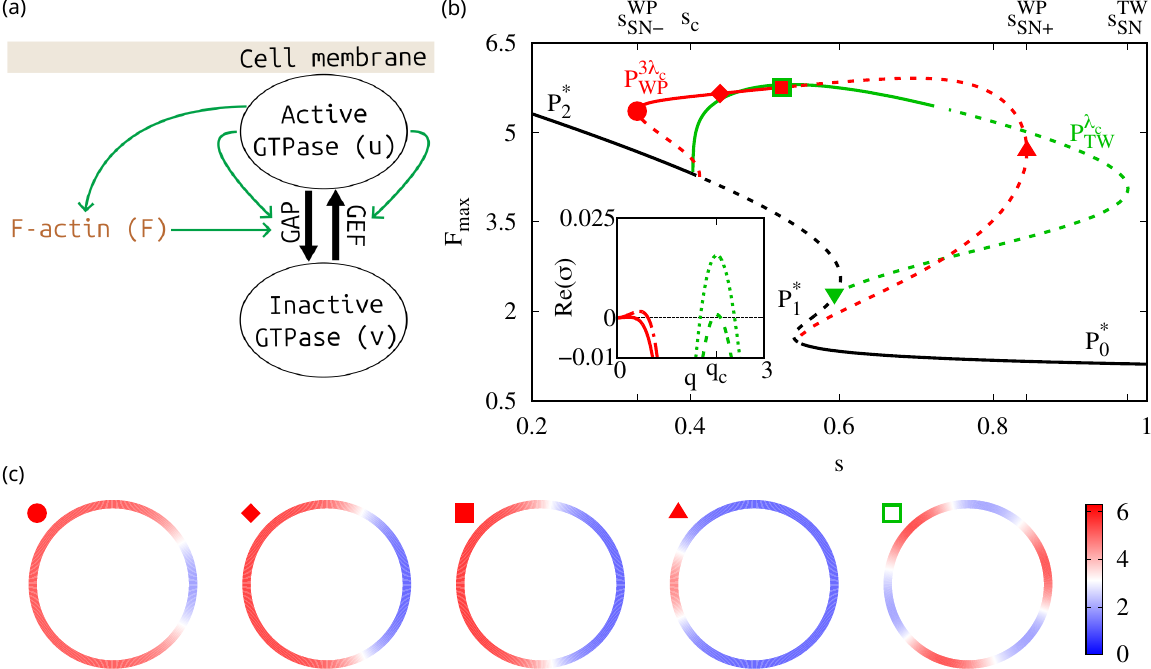}
		\caption{(a) Schematic representation of model system~\ref{eq:model}. (b) Bifurcation diagram showing the uniform steady states $\pb_{0,1,2}^\ast$ (black), polarized mesas $\pb^{3\lambda_c}_{\rm WP}$ (red), and traveling waves $\pb^{\lambda_c}_{\rm TW}$ (green), where superscripts denote the domain lengths and $\lambda_c\approx 3.093$. Solid lines indicate linear stability and dashed instability. The inset shows the dispersion relations of the codimension-2 instability onset at $s=s_c\approx 0.409$ and $q_c\approx 2.031$, and after the instability at $s\approx0.417$; the solid (dashed-dotted) line refers to the real values while the dashed (dotted) are complex conjugates. (c) Profiles of mesa solutions on a ring at selected locations along $\pb^{\lambda_c}_{\rm TW}$ according to the respective symbols, $s:({\color{red}{\scalebox{1.5}{$\bullet$}}},{\color{red} \blacklozenge},{\color{red}\blacksquare},{\color{red}\blacktriangle},{\color{Green}\square})\approx(0.337,0.445,0.525,0.844,0.525)$, and `${\color{Green}\blacktriangledown}$' at $s\approx 0.593$ marks the parity breaking bifurcation onset of $\pb^{\lambda_c}_{\rm TW}$ associated with the $\pb^{\lambda_c}_{\rm WP}$ branch (not shown here). Other parameters: : $b=b_c\approx 0.067$, $\gamma=3.557$, $\eta=0.6$, $p_0=0.8$, $p_1=3.8$, $D=0.1$, $D_F=0.001$, and $M=2$.}
		\label{fig:math}
	\end{figure*}
	%***********************************************************************
	
	One of the common model organisms to study basic mechanisms of switching between different migratory modes is the social amoeba \textit{Dictyostelium discoideum} (\textit{D.~discoideum}).
	Besides their common amoeboid motility, \textit{D.~discoideum} cells may also exhibit a keratocyte-like (so-called ``fan-shaped'') mode of locomotion. While amoeboid motility is characterized by the formation of highly dynamic, localized pseudopodia, resulting in erratic, random displacements, fan-shaped cells move in a highly persistent fashion and show a stable cell shape that is elongated perpendicular to their direction of motion~\cite{asano_keratocyte-like_2004}. Fan-shaped \textit{D.~discoideum} cells were observed as a consequence of genetic mutations~\cite{asano_keratocyte-like_2004,moreno2020modeling} and specific developmental conditions~\cite{cao2019plasticity}.
	Their persistent forward motion is driven by a ring-shaped actin wave that covers most of their ventral membrane~\cite{asano_correlated_2008}. Actin waves of this type have been thoroughly characterized in \textit{D.~discoideum}~\cite{gerisch_mobile_2004,gerisch2011different,gerhardt_actin_2014}.
	They serve as precursors of endocytic cups and their formation is controlled by small GTPase signaling~\cite{gerisch_self-organizing_2009,veltman_plasma_2016}.
	Thus, the fan-shaped mode was also directly induced in \textit{D.~discoideum} cells by synthetically increasing their level of small GTPase activity~\cite{miao_altering_2017,iwamoto2025excitable}. However, spontaneous switching within the same cell was also observed~\cite{Moldenhawer2022Switching}.
	
	The above examples of switching between distinct motility modes can be explained in different ways, e.g. by invoking changes in connectivity of the underlying regulatory network, or by changes in system parameters. In this study, we focus on a third possibility, where distinct motility modes coexist in the same regulatory systems and for a fixed set of parameters, so that external perturbations merely trigger switches between them.
	We first devise and analyze a simplified reaction-diffusion model with mass conservation, mimicking the coupled GTPase and F-actin dynamics to reveal a bistability region, where polarized and oscillatory states generically coexist. 
	Secondly, we demonstrate why, upon a large perturbation, the polarized state may become unstable, resulting in a transition to oscillatory dynamics, associated with transitions between two coexisting migratory modes.
	Finally, we show an experimental demonstration of a fan-shaped \textit{D.~discoideum} cell that, upon collision with a rigid boundary, becomes unstable and undergoes a transition to a disordered non-polar mode of migration.
	\newline \newline
	\textit{Model equations, bifurcation analysis, and coexistence --}
	Actin waves appear to be ubiquitous in eukaryotic cells~\cite{Allard2013,beta2017intracellular,beta2023actin}. In some species and under certain conditions, dynamic traveling wave (TW) patterns of F-actin and its regulators appear to coexist with static structures that are consistent with polarized cell fronts or cell division rings~\cite{inagaki2017actin,beta2023actin}.
	This motivated us to study whether a prototypical reaction-diffusion model of GTPase signaling coupled to F-actin can account for coexisting modes of cell migration.
	The model includes a slow diffusing active form ($u$) and a fast diffusing inactive form ($v$) of the GTPase (e.g. Rac or Ras) coupled to F-actin ($F$):
	\begin{subequations}\label{eq:model}
		\begin{align} 
			\partial_t u&= (b+\gamma u^2)v - (1+s F+u^2) u + D\nabla^2 u,\label{eq:model u}\\
			\partial_t v&= -(b+\gamma u^2)v+ (1+s F+u^2) u  + \nabla^2 v,\label{eq:model v}\\
			\partial_t F&=\eta (p_0+p_1 u - F) +D_F \nabla^2 F.\label{eq:model F}   
		\end{align}
	\end{subequations}
	System~\eqref{eq:model} is a simplified version of the model of Holmes \textit{et al.}~\cite{holmes2012regimes} for actin waves.
	The model structure is supported by later experiments in Xenopus oocyt showing that the GTPase Rho self-activates via Ect2, and F-actin inactivates it via RGA-3/4~\cite{Michaud2022}.
	
	The $(u,v)$-subsystem is of a gradient nature and steady states of it are obtained through the Maxwell construction~\cite{verschueren2017model,brauns2020phase,champneys2021bistability}. Nevertheless, it acts as the ``bistable" part, and slow negative feedback is provided by F-actin ($F$), whose role is similar to a ``refractory variable'' in the FitzHugh-Nagumo (FHN) model~\cite{fitzhugh1961impulses,nagumo1962active}, see also the schematic representation in Fig.~\ref{fig:math}(a). However, unlike the dissipative models, system~\eqref{eq:model} conserves the total amount of \mbox{GTPase},
	\[
	M:=\Omega^{-1}\int_\Omega\left[u+v\right]{\rm d}\mathbf{x}=\text{constant},
	\]
	where $\Omega$ is the integration domain. As we are primarily interested in the dynamics at the cell edge, we consider a one-dimensional domain (1D) with periodic boundary conditions (i.e., a ring geometry).
	
	An important feature in~\eqref{eq:model} is the basal rate of activation $b>0$; mathematically, this term excludes the existence of trivial solutions. In addition, $\gamma$ is the rate of auto-activation (positive feedback of active GTPase to its own activation rate), $s$ is the F-actin dependent inactivation rate, $\eta$ is the F-actin time scale parameter, $p_0$ is the F-actin basal growth rate, $p_1$ is the GTPase dependent F-actin assembly rate, and $D_F\ll D<1$ are the diffusion coefficients of F-actin and active GTPase, respectively. In what follows, we use $s$ as a control parameter while keeping all other parameters fixed. We note that a more detailed mathematical analysis of~\eqref{eq:model} is presented in~\cite{hughes2024travelling}.
	
	We start by numerically computing the uniform steady states $\pb^\ast=(u^\ast,v^\ast,F^\ast)$ of~\eqref{eq:model}, which result in up to three biologically relevant solutions, $\pb^\ast_{0,1,2}>0$, forming an inverse ``S'' form with $\pb^\ast_0$, as shown in Fig.~\ref{fig:math}(b). Linear stability analysis in 1D of $\pb^\ast$ to infinitesimal perturbations~\cite{cross1993pattern}, leads to solutions $\pb(x,t)-\pb^\ast\propto \exp(\sigma t+iqx)$, where $\sigma$ is the growth rate of wavenumber $q$, resulting in three dispersion relations for $\sigma(q;s)$. We find that, while $\pb^\ast_0$ is linearly stable, the state $\pb^\ast_2$ exhibits a simultaneous long-wavelength and finite wavenumber Hopf instability at $(s,b)=(s_c,b_c)\approx(0.409,0.067)$, as shown in the inset of Fig.~\ref{fig:math}(b); such simultaneous instabilities are also known as codimension-2 bifurcations, cf.~\cite{Yochelis2022}. The former instability occurs around $q=0$ and leads to the bifurcation of steady states in subcritical direction (towards a stable portion of $\pb^\ast_2$, $s<s_c$). The latter gives rise to both traveling and standing waves (TWs and SWs with $q_c\approx 2.031$, respectively) that bifurcate supercritically (towards the unstable portion of $\pb^\ast_2$, $s>s_c$) and where TWs are linearly stable (here, SWs are ignored and for further details we refer to~\cite{hughes2024travelling}). We used the package AUTO~\cite{auto} to compute the bifurcating branches and solutions; linear stability of nonuniform states is obtained via the \textit{eigs} function in MATLAB.
	
	In our context of cells, whose size is finite, we set the domain length $L=3\lambda_c$, where $\lambda_c=2\pi/q_c \approx 3.093$ and $q_c\approx 2.031$ is the critical wavenumber at the onset of the finite wavenumber Hopf instability ($s=s_c$).
	In Fig.~\ref{fig:math}(b) we show that the WP solutions, $\pb^{3\lambda_c}_{\rm WP}$, bifurcate from $s\gtrsim s_c$ and fold to the right (at $s=s_{\rm SN-}^{\rm WP}$), where their profile resembles a hole-like state, as shown in Fig.~\ref{fig:math}(c). Then, the branch continues to the right, folds again to the left (at $s=s_{\rm SN+}^{\rm WP}$), and terminates near the fold of $\pb^\ast_0$ at an additional long-wavelength instability. At $s=s_{\rm SN+}^{\rm WP}$, the profile resembles a peak-like solution, whereas between the folds, the profiles correspond to \textit{mesa} states~\cite{elliott1986mesa,kolokolnikov2006mesa}, which are also known as wave-pinning solutions~\cite{mori2011asymptotic,holmes2012comparison,verschueren2017model}, as shown in Fig.~\ref{fig:math}(c) by selected intermediate profiles.
	%***********************************************************************
	\begin{figure}[tp]
		{\includegraphics[width=0.48\textwidth,valign=t]{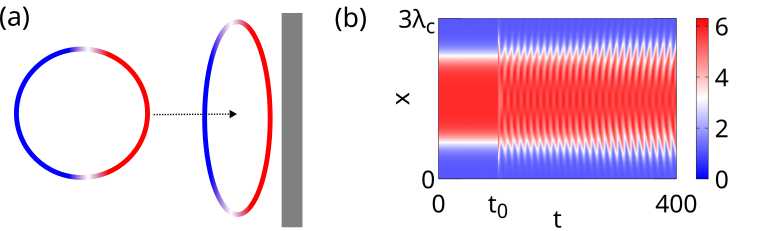}}
		(c){\includegraphics[width=0.45\textwidth]{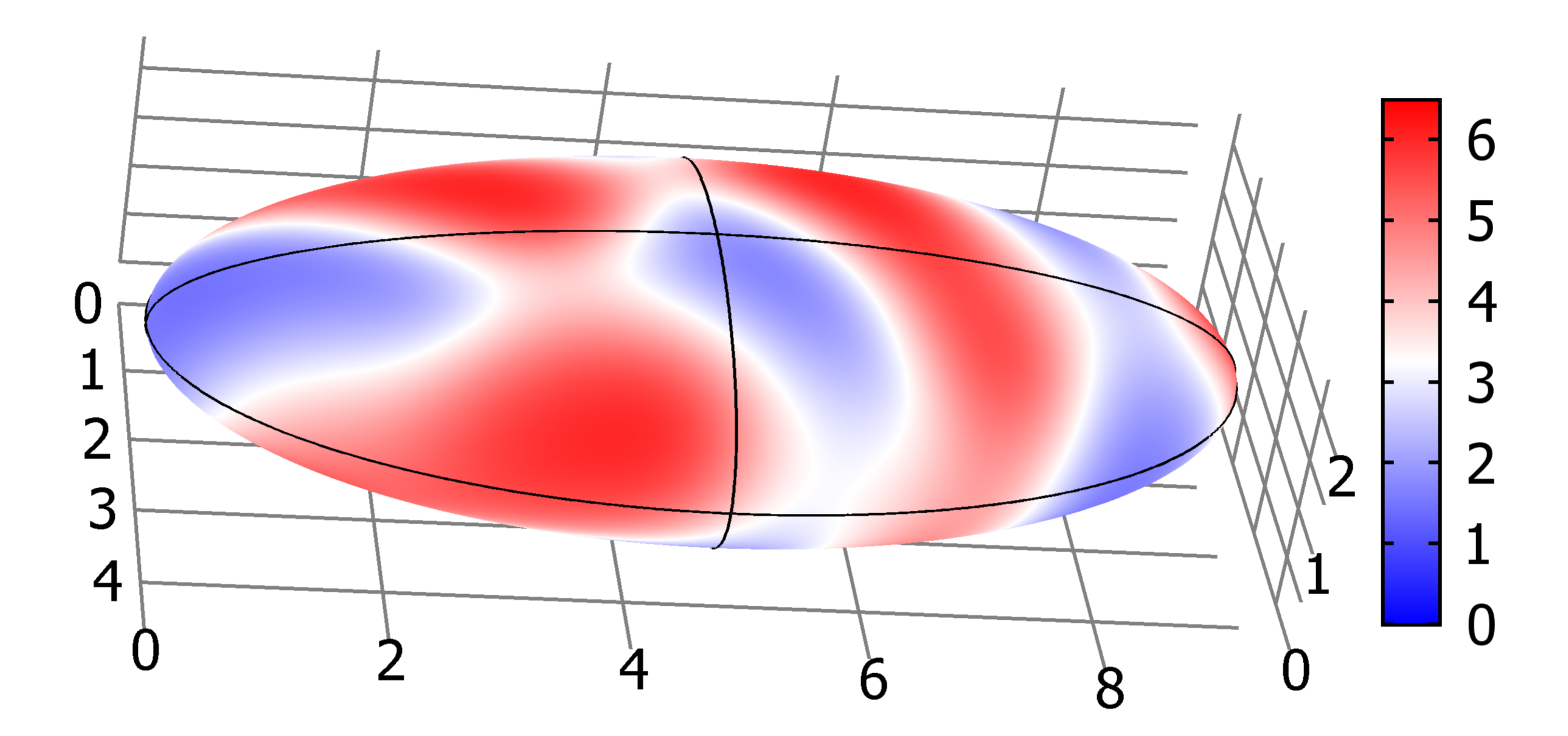}}
		\caption{(a) Schematic representation of a collision, emphasizing the possible widening of the mesa state. (b) Direct numerical integration of~\eqref{eq:model} with periodic boundary conditions on a domain length $L=3\lambda_c$ at $s=0.525$, showing the evolution of a mesa state into oscillations upon a perturbation applied at $t_0=100$. (c) Numerical integration as in (b) but on an ellipsoid surface geometry (using COMSOL Multiphysics\textsuperscript{\textregistered}) with principle diameters $d_x=3\lambda_c\approx 9.28$, $d_y=3\lambda_c/2\approx 4.64$, $d_z=3\lambda_c/4 \approx 2.32$, and random initial conditions. Here, we show a snapshot at $t=100$. (See SM movie.) Other parameters as in Fig.~\ref{fig:math}.}
		\label{fig:pert}
	\end{figure}
	%***********************************************************************
	%
	%***********************************************************************
	\begin{figure*}[tp]
		%\centering
		{\includegraphics[width=0.9\textwidth]{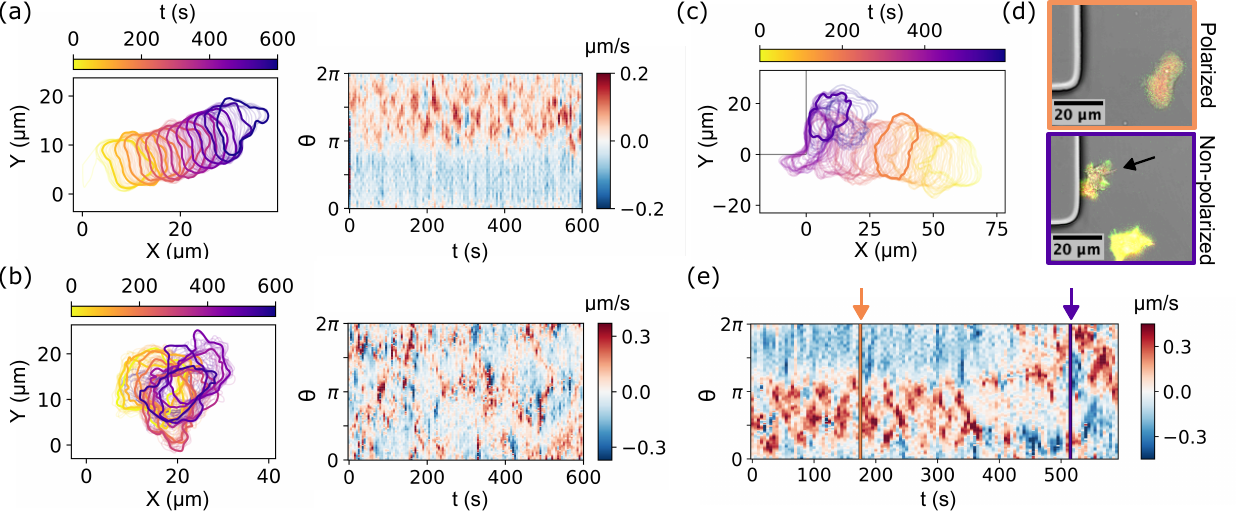}}
		\caption{Migratory modes and switching between them in \textit{D.~discoideum} after wall collision. (a)~Cell contour series and corresponding local motion kymograph for a polarized fan-shaped cell. (b)~Same as (a) for a non-polarized cell. (c)~Cell contour series for a cell colliding with a barrier, where it switches from polarized to non-polarized migration upon collision, see also SM movie. (d)~Confocal microscopy images of the cell shown in (c), where the cell expresses LifeAct-GFP and PH$_{\mathrm{CRAC}}$-mCherry. The arrow in the non-polarized image indicates the cell of interest. (e)~Local motion kymograph from the contours in (c), where the vertical orange (resp. purple) line and arrow indicate where the snapshot of a polarized (resp. non-polarized) cell was taken. The local motion kymographs show the displacements of equidistant points along the cell contour obtained by minimizing the sum of the squared displacements of all points, for details see~\cite{schindler_analysis_2021}.}
		\label{fig:exp}
	\end{figure*}
	%***********************************************************************
	
	To compute the bifurcating TW branch, we employ a comoving frame $\xi=x-ct$, where $c$ is the speed and at the onset ($s=s_c$) is given by the phase speed $c_c={\rm Im}\, \sigma/q_c\approx 0.191$. The TWs branch, $\pb^{\lambda_c}_{\rm TW}$, is supercritical, i.e., in the direction of increasing $s$ and after a fold at $s_{\rm SN}^{\rm TW}$, it ends in a parity-breaking bifurcation on the $\pb_{\rm WP}$ branch with the corresponding length $L=\lambda_c$ [`${\color{Green}\blacktriangledown}$' in Fig.~\ref{fig:math}(b)], for details see~\cite{hughes2024travelling}. Figure~\ref{fig:math}(b) shows the coexisting branch of TWs and, moreover, the bistability region, where the steady mesa states and TWs are linearly stable. Next, we address the role of large amplitude perturbations within this bistable regime, as in the case of collisions.
	\newline \newline
	\textit{Bistability and the role of large perturbations --} In the case of motile cells, the steady mesa states correspond to strongly polarized F-actin distributions and, therefore, can be taken as representations of fan-shaped cells with highly polarized, directed migration. Let us consider a polarized cell moving towards a wall and, upon collision, extending laterally, as schematically shown in Fig.~\ref{fig:pert}(a). One of the possible and simplest nonlinear perturbations of the mesa state is that it may become wider. However, due to mass conservation (here $M=2$), the mesa state cannot maintain its amplitude, i.e., its peak value will decrease at the expense of widening~\cite{maree2012cells}. To ensure the mass constraint, we first fixed the value of $s$ for the ``pre-perturbed state'' on the stable portion of $\pb^{3\lambda_c}_{\rm WP}$ [`${\color{red}\blacksquare}$' in Fig.~\ref{fig:math}(c)]. Then, we select a state with a different value of $s$ [`${\color{red} \blacklozenge}$' in Fig.~\ref{fig:math}(c)], representing a 21\% increase in the width of the top plateau (for the same $M$ value). We note that collisions comprise many other perturbations (e.g., parameter variations in space~\cite{Goryachev2016,noguchi2025nonequilibrium}) that are likely to enhance the instability of the polarized state (mesa states). However, such perturbations are model-dependent and, thus, beyond the scope of our simplified framework Eqs.~\ref{eq:model}.
	
	Solving Eqs.~\eqref{eq:model} via direct numerical integration on domain length $L=3\lambda_c$, we show that the "perturbed" initial state develops oscillations leading to outward propagating TWs, as shown in Fig.~\ref{fig:pert}(b). However, the form of the oscillations is of lesser importance, as they lead to disordered dynamics in higher spatial dimensions due to modulational instabilities~\cite{Cross1993,pismen2006patterns}, even in the absence of stochastic contributions or shape deformations~\cite{machacek2006morphodynamic,Weiner2007,dreher2014spiral,cao2019plasticity,moreno2020modeling,liu2021spots,hladyshau2021spatiotemporal,Michaud2022,TeamKeshet2024}.
	As an example, a snapshot of the resulting dynamics on an ellipsoid surface is displayed in Fig.~\ref{fig:pert}(c). The differences in colors correspond to different actin concentrations, representing intracellular forces that drive the formation of membrane protrusions and, thus, mimic the fluctuating directions of motion in the non-polarized mode of motility, see Fig.~\ref{fig:exp}. This means that, in general, oscillations and disordered patterns belong to the same universality class~\cite{holmes2017mathematical,cao2019plasticity,Liu2021,Michaud2022,iwamoto2025excitable}.
	\newline \newline
	\textit{Wall collisions of fan-shaped cells --}
	Inspired by our theoretical findings, we sought a simple experimental example of coexisting modes of migration, and how cells switch between them. Here we show that directed and irregular migratory modes of \textit{D.~discoideum} before and after a wall collision provide such an example. We recorded the migration of \textit{D.~discoideum} cells in PDMS-based microfluidic chambers of different geometry. For our imaging experiments we used a non-axenic knockout cell line (DdB wildtype background) that exhibited increased small GTPase activity~\cite{bloomfield_neurofibromin_2015}, resulting in the abundant formation of actin waves and an increased ratio of fan-shaped cells~\cite{veltman_plasma_2016,flemming_how_2020,Moldenhawer2022Switching}. The cell line furthermore expressed green and red fluorescent fusion proteins (LifeAct-GFP and PH$_{\mathrm{CRAC}}$-mCherry) allowing for fluorescence imaging of actin and PIP$_3$ dynamics, respectively, for details see SM.
	
	Our recordings showed repeated interactions of migrating cells with the side walls of the microfluidic chamber.
	While in most cases cells maintained their mode of migration, we also observed instances, where, upon hitting the boundary, a fan shaped cell lost its polarity and switched to a disordered mode of locomotion based on small and dynamically changing pseudopodia.
	An example is displayed in Fig.~\ref{fig:exp}, where a fan-shaped cell approaches a corner in the PDMS side wall of the chamber from the bottom right. The transition in the migratory mode upon collision with the side wall is reflected in a change of cell shape from the stable elongated fan to an irregular morphology with small protrusions around the cell border. 
	A color coded temporal sequence of cell contours before, during, and after the collision can be seen in Fig.~\ref{fig:exp}a, where examples of a polarized cell before the collision and a non-polarized cell after the collision are highlighted as bold contours.
	They correspond to the fluorescence images displayed in the two panels in Fig.~\ref{fig:exp}b.
	The loss of polarity upon collision with the side wall is also illustrated in a kymograph representation of the local motion of the cell border, shown in Fig.~\ref{fig:exp}c.
	During fan-shaped motion, a stable protruding cell front and a retracting back (0 to $\pi$ and $\pi$ to $2\pi$, respectively) can be seen in the kymograph until the collision at around $t=400$~s.
	After the collision, polarity is lost and protrusive and retractive activities are distributed all around the cell border.
	\newline \newline
	\textit{Discussion --} We uncover a mechanism for the transition from polarized to oscillatory dynamics, following the analysis of a mass conserving reaction-diffusion system that agrees with experimental recordings of collisions of \textit{D.~discoideum} cells with solid boundaries (Fig.~\ref{fig:exp}). In this mechanism, mass conservation is a key component, as similar behavior cannot arise in reaction-diffusion models without this feature (Fig.~\ref{fig:math}). While polarized F-actin distributions are associated with fan-shaped motility, the oscillatory states represent F-actin dynamics that result in non-polarized disordered motility (Fig.~\ref{fig:pert}).
	Moreover, our theory is also consistent with so-called ``crawling'' and ``ruffling'', corresponding to TWs that move along the edge of adherent cells~\cite{dobereiner2006lateral,giannone2004periodic,yang2019two}.
	Taken together, our results suggest that GTPase signaling coupled to F-actin feedback~\cite{landino2021rho,bement2024patterning,iwamoto2025excitable} may account for coexisting migratory modes in eukaryotic cells, an essential prerequisite to understanding and potentially controlling their motility, paving the way toward novel functionalities and applications. 
	\newline
	
	\begin{acknowledgments}
		\noindent This work was funded by the Deutsche Forschungsgemeinschaft (DFG, project-ID No. 318763901–SFB1294), Germany (C.M.T. and C.B.), the Natural Sciences and Engineering Research Council of Canada (NSERC) CGS-D Scholarship (J.M.H.), the NSERC Discovery Grant (L.E.K.), and the United States - Israel Binational Science Foundation (BSF, grant no. 2022072), Jerusalem, Israel (A.Y.).
	\end{acknowledgments}
	
	\section*{AUTHOR DECLARATIONS}
	\noindent \textbf{Conflicts of Interest.} The authors have no conflicts to disclose.
	
	\section*{DATA AVAILABILITY}
	\noindent The data that support the findings of this study are available from the corresponding author upon reasonable request.
	
	\section*{REFERENCES}
	%\bibliography{refs}

	%\newpage
	\renewcommand{\thesection}{S\arabic{section}} 
	\renewcommand{\thepage}{S\arabic{page}}
	%\renewcommand{\theequation}{S\arabic{equation}}
	
	%\linespread{1.5}
	\section*{Supplementary Material}
	
	\section{Cell culture}%\label{sec:exp_S4}
	\noindent
	Non-axenic \textit{D.~discoideum} cells (DdB wildtype background) that are deficient in NF1~\cite{bloomfield_neurofibromin_2015} were transformed with an episomal plasmid encoding Lifeact-GFP and PHcrac-RFP as described in~\cite{flemming_how_2020}. Cells were cultivated in 10~cm dishes with Sørensen's buffer supplemented with 50~µM MgCl$_2$ and 50~µM CaCl$_2$ (Sørensen's-MC buffer) and using G418 (5~µg/ml) and hygromycin (33~µg/ml) as selection markers. \textit{Klebsiella aerogenes} with an OD600 of 20 were added to the solution in 1:10 volume to a final OD600 of 2. Prior to imaging, remaining bacteria were removed by two-times centrifugation at $300\,\times$~g, and reconstituting the resulting pellet in Sørensen's-MC buffer. To increase the number of fan-shaped cells, cells were starved for 1 hour before infusion in the microfluidic chip.
	
	\section{Microfluidic chips}
	\noindent
	Direct write lithography with a maskless aligner (µMLA, Heidelberg Instruments Mikrotechnik GmbH, Germany) was used to pattern a silicon wafer coated with a 10~µm photoresist layer (SU-8 2010, Micro Resist Technology GmbH, Germany). Polydimethylsiloxane (PDMS, Sylgard 184, Dow Corning GmbH, Germany) at a ratio of 10:1 (base to curing agent) was poured onto the microstructured wafer, degassed and cured for 2~h at 75°C. The microfluidic chips were assembled by plasma bonding the PDMS blocks to a glass coverslip (\#1.5, $24\times 40$~mm, Menzel Glaser). The chips were immediately filled with Sørensen's buffer (14.7~mM KH$_2$PO$_4$, 2~mM Na$_2$HPO$_4$, pH~6.0) supplemented with 50~µM MgCl$_2$ and 50~µM CaCl$_2$ and rinsed extensively before adding the cell solution.
	
	\section{Imaging and data processing}
	\noindent
	Imaging was performed using a laser scanning microscope (LSM780, Zeiss, Jena) with a 488~nm Argon laser and a 561~nm diode laser, using a $40\times$ oil immersion objective. Timelapse recordings at an interval of 5~s were acquired within less than 4~hours of starvation, and without any external flow in the microfluidic chip. Cell contours were obtained using the RFP channel as follows: fluorescent images were filtered using a median filter (2 pixel size) in Fiji~\cite{Schindelin2012}, and used as input for the `find contours` function of the scikit-image python package ~\cite{walt_scikit-image_2014}. Only the contours corresponding to the cell outlines were kept and processed with AmoePy~\cite{schindler_analysis_2021,schindler_amoepy_2025} to obtain local motion kymographs. 
	
	%\section*{REFERENCES}
	%\bibliography{refs}
	
\end{document}